\documentclass[runningheads]{llncs}
\usepackage{graphicx}
\usepackage{hyperref}
\usepackage{multirow}
\usepackage{listings}
\usepackage{wrapfig}
\usepackage{lscape}
\usepackage{rotating}

\usepackage[many]{tcolorbox} 
\newcommand{\daniel}[1] {\textbf{\textcolor{red}{~#1}}}

\newcommand{\sheeba}[1] {\textbf{\textcolor{blue}{~#1}}}
\begin{document}
\title{FAIR Jupyter: a knowledge graph approach to semantic sharing and granular exploration of a computational notebook reproducibility dataset}
\titlerunning{FAIR Jupyter}
%
\author{Sheeba Samuel\inst{1}\orcidID{0000-0002-7981-8504} \and
Daniel Mietchen\inst{2,3,4}\orcidID{0000-0001-9488-1870} }
\authorrunning{Samuel et al.}
%

\institute{Chemnitz University of Technology, Germany 
\email{sheeba.samuel@informatik.tu-chemnitz.de}
\and FIZ Karlsruhe — Leibniz Institute for Information Infrastructure, Germany \and 
Ronin Institute, Montclair, New Jersey, United States \and 
Institute for Globally Distributed Open Research and Education (IGDORE)
\email{daniel.mietchen@fiz-karlsruhe.de}
}
%

\maketitle              
\begin{abstract}
The way in which data are shared can affect their utility and reusability.
Here, we demonstrate how data that we had previously shared in bulk can be mobilized 
further 
through a knowledge graph that allows for much more granular exploration and interrogation. 
The 
original
dataset is about the computational reproducibility of GitHub-hosted Jupyter notebooks associated with biomedical publications.
It contains rich metadata about the publications, associated GitHub repositories and Jupyter notebooks, and the notebooks' reproducibility. 
We took this dataset,
converted it into semantic triples and loaded these into a triple store 
to create a knowledge graph~-- 
FAIR Jupyter~-- 
that we made accessible via a web service.
This 
enables granular data exploration and 
analysis
through queries
that
can be tailored to specific use cases.
Such queries
may
provide details about 
any of the variables from the original dataset,
highlight relationships between 
them
or 
combine 
some of 
the graph's content with
materials from corresponding external resources.
We provide a collection of example queries
addressing
a range of 
use cases 
in research and education.
We also outline how
sets of such queries can be used to profile specific content types, either individually or by class.
We conclude by
discussing
how such a semantically enhanced sharing of complex datasets can
both
enhance their FAIRness~-- i.e.,\ their findability, accessibility, interoperability, and reusability~-- 
and help
identify and communicate best practices, 
particularly with regards to data quality, standardization, automation and reproducibility.




\end{abstract}

\newpage


\section{Introduction}
\label{sec:Introduction}

Data sharing has many facets, including the nature of the data, the purpose of the sharing, reuse considerations as well as various practicalities like the choice of file formats, metadata standards, licensing and location for the data to be shared.

Here, we 
look into some of the practical and reuse aspects by 
%
mobilizing a
previously shared reproducibility
dataset
in a more user-friendly fashion.
That previous dataset arose from a study \cite{samuel2024computational} 
of the computational reproducibility of Jupyter notebooks associated with biomedical publications.
It is already publicly available as a 1.5GB SQLite database contained within a ZIP archive of 415.6 MB (compressed) \cite{samuel2023Dataset}
but in order to be explored, it needs to be unzipped, loaded into a SQLite server and queried via SQL. 
While these steps are routine for many, they nonetheless present a technical hurdle that stands in the way of
broader use
of the data, both in research and educational contexts. 

For instance, imagine
an instructor of a programming course for 
wet lab
researchers 
who wants to present to her students some 
real-world
Jupyter notebooks 
from their respective research field 
that use a specific Python module and are either fully reproducible or exhibit a given type of error.
Wouldn't it be nice~-- and quicker than via the route outlined above~-- if she could get her students to run such queries directly in their browser, with no need to install anything on their system? 
Enabling students and instructors to do this~-- or indeed anyone else, from reproducibility researchers to journal editors or package maintainers~-- 
is what we are 
aiming at.

One way to build such a webservice would be to use a web framework like Flask or Django in combination with a library like sqlite3 \cite{sqlite}
to interact with the database. 
However, this approach complicates semantic integration with other resources, and so we chose instead to 
leverage semantic web standards \cite{pan2009resource}
and build a
demonstrator for converting a dataset 
into a knowledge graph (KG).


While we are not aware of knowledge graphs about Jupyter notebooks or GitHub repositories, there have been a number of related efforts. These include
the
creation of 
knowledge graphs about 
FAIR computational workflows \cite{goble2020fair},
PubMed \cite{xu2020building}, 
scientific software \cite{kelley2021a} and artificial intelligence tasks and benchmarks \cite{blagec2022curated}, along with tools for creating
schemas and knowledge graphs from data
\cite{hubert2024pygraft}.
%
%
%
%
%
Of particular relevance here is the Open Research Knowledge Graph \cite{jaradeh2019open}, whose potential for assisting with reproducibility has been 
outlined
\cite{hussein2023increasing}.




\section{Methods}
\label{sec:Method}

\begin{figure}[h!]
\includegraphics[width=\textwidth]{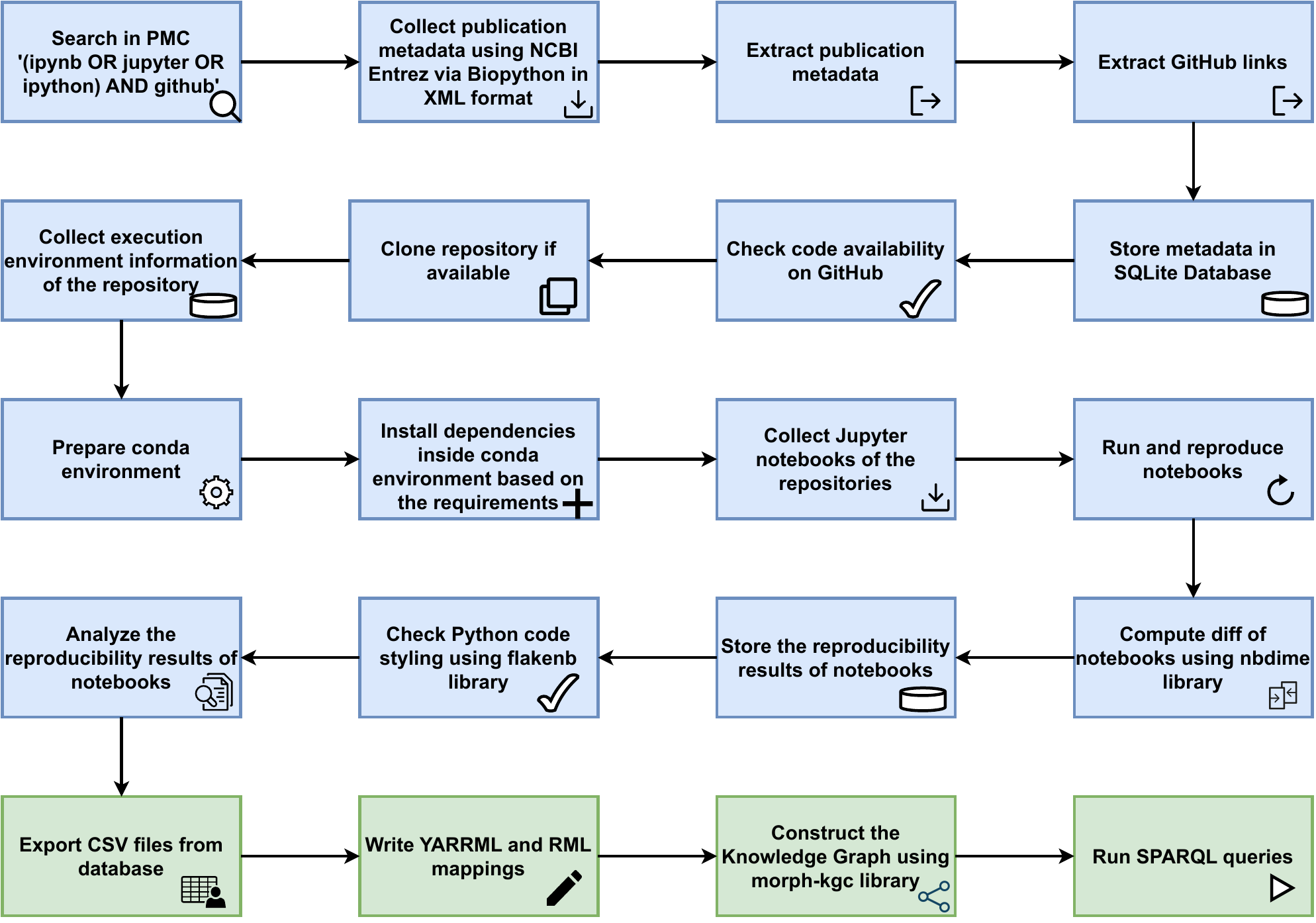}
\centering
\caption{Workflow overview. The blue workflow was used to construct the original dataset 
\cite{samuel2023Dataset}
and is described in \cite{samuel2024computational},
whereas
the subsequent knowledge graph construction
workflow shown in green represents the current study.
} 
\label{fig:KGmethodworkflow}
\end{figure}

Our workflow 
has two main components, as shown in Figure \ref{fig:KGmethodworkflow}. The first is the generation of the Jupyter notebook reproducibility dataset.
The second is 
the conversion of this dataset into 
the FAIR Jupyter knowledge graph, which forms the focus of the present study.


\subsection{Computational Reproducibility Dataset Generation}
This section represents a summary of the methodology used in \cite{samuel2024computational}. Briefly, we queried PubMed Central (PMC, cf.\ \cite{roberts2001pubmed})
for publications mentioning GitHub alongside keywords such as ``Jupyter'', ``ipynb'' (the file extension for Jupyter notebooks), or ``IPython'' (a predecessor to Jupyter).\footnote{We used the search query: ``(ipynb OR jupyter OR ipython) AND github''.}\label{foot:search-query}
 Utilizing the primary PMC IDs obtained this way, we then retrieved publication records in XML format using the \textit{efetch} function and collected publication metadata from PMC using NCBI Entrez utilities via Biopython \cite{cock2009biopython}.

Next, we processed the XML data retrieved from PMC by storing it in an SQLite database. Our database encompassed details regarding journals and articles, populating it with metadata including ISSN (International Standard Serial Number), journal and article titles, PubMed IDs, PMC IDs, DOIs, subjects, submission, acceptance, and publication dates, licensing information, copyright statements, keywords, and GitHub repository references mentioned in the publication. 
Additionally, we extracted associated Medical Subject Headings (MeSH terms) \cite{mesh} for each article. These terms, assigned during indexing in the PubMed database, are hierarchical. We obtained the top-level MeSH term by querying the MeSH RDF API through SPARQL queries to the SPARQL endpoint \cite{meshsparql}. This aggregation resulted in 108 top-level MeSH terms in our dataset, serving as proxies for the subject areas of the articles.

We extracted the GitHub repositories  mentioned in each article, including the abstract, body, data availability statement, and supplementary sections.\footnote{Since the GitHub repositories had been stated in a number of different formats, we harmonized them to ``https://github.com/{username}/{repositoryname}''.}
After this preprocessing, we associated each article with the GitHub repositories  extracted from it as well as with 
the journal in which the article had been published,
and we gathered author information in a separate database table, including first and last names, ORCID, and email addresses.

We verified the availability of GitHub repositories mentioned in the articles and, if existing, cloned them, based on the main branch, and gathered repository details including creation, update, and push dates, releases, issues, license details, etc.  using the GitHub REST API \cite{githubrestapi}. 
Additionally, we extracted details for each notebook provided in the repository, such as name, nbformat, kernel, language, cell types, and maximum execution count, and extracted source code and output from each cell using Python Abstract Syntax Tree (AST) for further analysis.

After the notebook collection, we gathered execution environment details by examining dependency declarations in repository files like requirements.txt, setup.py, and pipfile. After collecting necessary Python notebook execution information, we prepared a conda environment based on the declared Python version, installing dependencies listed in files such as requirements.txt, setup.py, and pipfile. For repositories lacking specified dependencies, the pipeline executed notebooks by installing all Anaconda dependencies, leveraging Anaconda's comprehensive data science package suite.
We also conducted Python code styling checks using the flakenb \cite{flake8nb} library, which enforces code style guidelines outlined in PEP 8, to collect all detected errors, obtaining information on the error code and description.

We executed our pipeline on 27\textsuperscript{th} March 2023 which ran till 9 Mai 2023, for a total of 43 days. The code has been adapted from \cite{pimentel2019a,samuel2021reproducemegit}. We utilize this method to reproduce Jupyter notebooks from GitHub repositories, as outlined in \cite{pimentel2019a}. Additionally, we leverage ReproduceMeGit \cite{samuel2021reproducemegit}, which uses the nbdime library \cite{nbdime} to compare execution results with the original results. This forms the foundation of our code for the reproducibility study.

\subsection{FAIR Jupyter KG Construction}
\subsubsection{Data modeling}
\begin{figure}
\includegraphics[width=\textwidth]{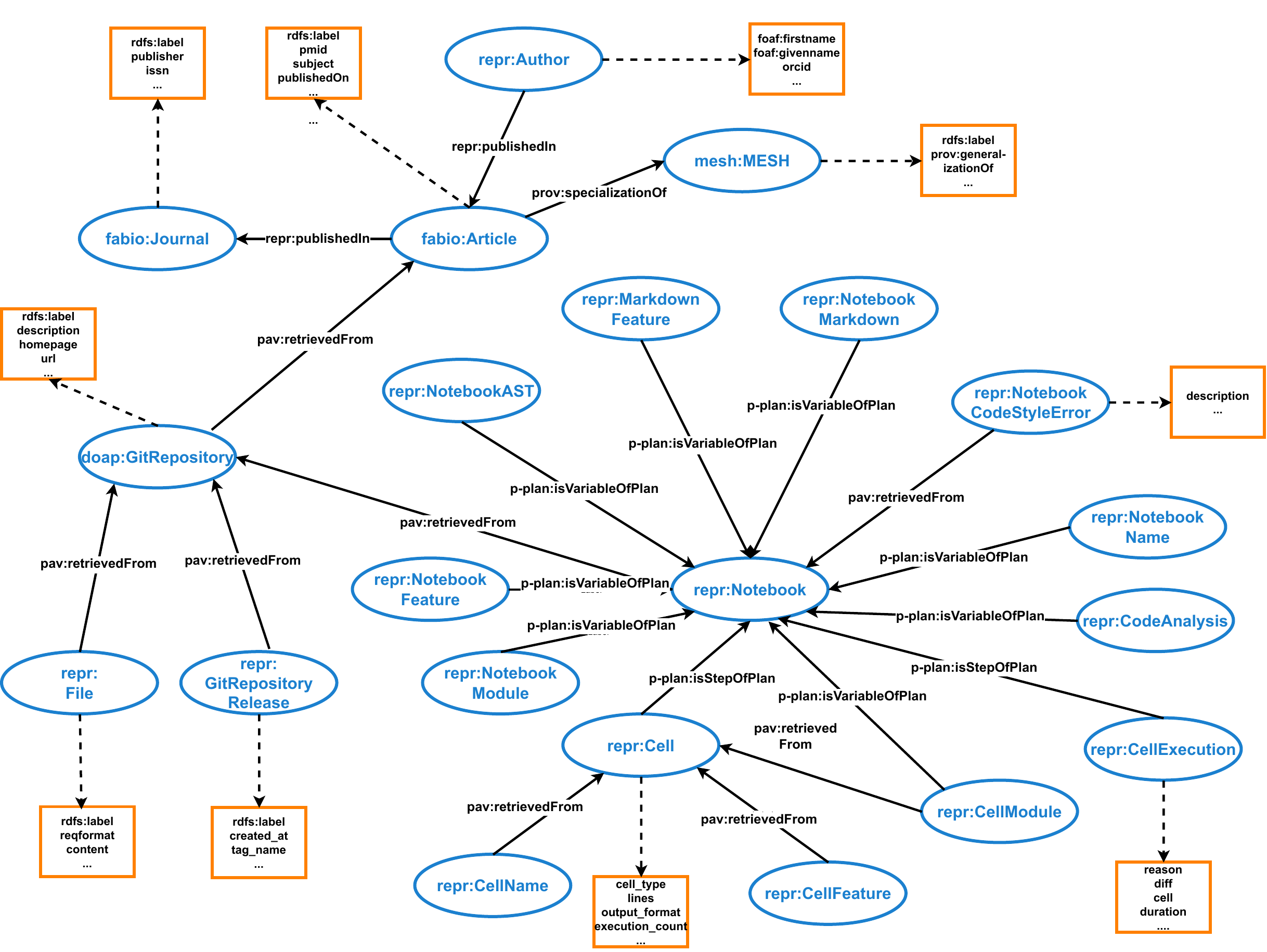}
\centering
\caption{Partial outline of the data model used in FAIR Jupyter. Classes of entities (represented by ellipses) and the class properties (represented by orange rectangles)
were inferred from the original dataset, and~-- along with relationships between them (arrows)~-- expressed in terms of relevant ontologies.
Note that  
requirement files and repository files are both represented as \texttt{repr:File}.
}.

\label{fig:model}
\end{figure}

In this section, we provide a brief description of the ontology model 
used in the construction of the FAIR Jupyter knowledge graph.
Overall, it contains 22 classes, as outlined in Figure \ref{fig:model}. They are centred around notebooks, notebook cells, repositories and articles, each of which are linked to several other classes.
We reused the following ontologies for 
describing these entities: 
PROV-O \cite{lebo2013prov}, REPRODUCE-ME \cite{samuel2018combining}, P-Plan \cite{garijo2012augmenting}, PAV \cite{ciccarese2013pav}, DOAP \cite{doap} and FaBiO \cite{peroni2012fabio}.

Building on PROV and P-Plan, the REPRODUCE-ME ontology captures provenance information for individual Jupyter Notebook cells \cite{samuel2018combining}, in addition to the end-to-end scientific experiment with real-life entities like instruments and specimens, as well as human activities such as lab protocols and screening \cite{samuel2022end}. Hence, we use and extend this model to construct the FAIR Jupyter KG.

We reused the \texttt{fabio:Article} to link the publications in our dataset and  \texttt{fabio:Journal} to represent the journal where the article was published. 
FaBiO is an ontology that helps represent information about publishable works (articles, books, etc.) and the bibliographic references that connect them \cite{peroni2012fabio}.
The GitHub repository is represented using the class \texttt{doap:GitRepository} from the DOAP ontology \cite{doap}, which is used to describe open source software projects. We reused the class \texttt{repr:Notebook} to represent the Jupyter Notebooks, which is a subclass of \texttt{p-plan:Plan} extending the class \texttt{prov:Plan}.
The PROV-O \cite{lebo2013prov} is a W3C recommendation providing foundation to implement provenance applications. We use two object properties of PROV-O ontology: \texttt{prov:specializationOf} and \texttt{prov:generalizationOf} to show that the article is a specialization of a MESH term and the corresponding MESH term a generalisation of the top level MESH term in its hierarchy.
To denote where a resource is retrieved from (e.g., \texttt{repr:Notebook} is \texttt{pav:retrievedFrom} \texttt{doap:GitRepository}), we have used 
the object property \texttt{pav:retrievedFrom}. 
PAV which builds on the PROV-O ontology, is a lightweight ontology for tracking Provenance, Authoring and Versioning and describes information about authorship, curation, and versioning of online resources \cite{ciccarese2013pav}.
Each individual cell of a computational notebook \texttt{repr:Cell} is described as step of a \texttt{repr:Notebook} using the object property \texttt{p-plan:isStepOfPlan}. 
%
To describe the different features of notebooks and their cells, we reused the object property \texttt{p-plan:isVariableOfPlan}.

\subsubsection{Mappings and KG construction}
We used the csv files exported from the database as input for the YARRRML mapping \cite{heyvaert2018declarative}. 
YARRRML is a human-readable text-based format for expressing declarative rules to generate Linked Data from different data sources. Listing \ref{lst:yarrrmlmapping} shows an example of YARRRML mapping used for expressing the rules for `repositories' and `notebooks'.
To facilitate adaptation to a variety of use cases,
we created mappings for each entities in separate files. The mappings are created for the classes of the ontology  (see Figure \ref{fig:model}).
We chose the same name for the user-chosen keys as the name provided in the database tables. 
The key source has value of the corresponding csv file of the entity.
We used the namespace \url{https://w3id.org/reproduceme/} for generating the IRI.
We reused the properties from the existing ontologies wherever possible to generate the combination of predicates and objects.
For others, we added in the REPRODUCE-ME ontology.
The YARRRML mappings were written using the YARRRML editor, Matey \footnote{\url{https://rml.io/yarrrml/matey/}}. These mappings were then converted to RML mapping \cite{dimou2014rml}.
The generated RML mappings were further used as input to the Morph-kgc library \cite{arenas2024morph}.
Morph-KGC is a tool designed to build RDF knowledge graphs from different data sources using R2RML and RML mapping languages. For large datasets, Morph-KGc's use of mapping partitions significantly speeds up processing and reduces memory usage.
The triples generated were stored in N-triples format.

\begin{lstlisting}[caption={Part of a YARRRML mapping for the repositories and notebook
% \daniel{What is the tilde doing between the two "csv" strings?}\sheeba{That is the syntax of the mapping to show that the data format is csv}
},label={lst:yarrrmlmapping}]
mappings:
repositories:
    sources:
      - [data/repositories.csv~csv]
    s: https://w3id.org/reproduceme/repository_$(id)
    po:
      - [a, doap:GitRepository]
      - [rdfs:label,$(repository)]
      - p: pav:retrievedFrom
        o:
          - mapping: article
            condition:
              function: equal
              parameters:
                - [str1, $(article_id), s]
                - [str2, $(id), o]
  notebooks:
    sources:
      - [data/notebooks.csv~csv]
    s: https://w3id.org/reproduceme/notebook_$(id)
    po:
      - [a, repr:Notebook] 
      - [rdfs:label, $(name)]
      - [repr:kernel, $(kernel)]
      - [repr:language, $(language)]
      - p: pav:retrievedFrom
        o:
          - mapping: repositories
            condition:
              function: equal
              parameters:
                - [str1, $(repository_id), s]
                - [str2, $(id), o]
\end{lstlisting}

\subsubsection{Triple store and Web service}
We use Apache Jena Fuseki \cite{carroll2004jena} as a triple store to query our knowledge graph.
The FAIR Jupyter knowledge graph is available at 
\href{https://reproduceme.uni-jena.de/#/dataset/fairjupyter/query}{https://reproduceme.uni-jena.de/\#/dataset/fairjupyter/query}.

\newpage

\section{Results}
\label{sec:Results}
Table \ref{tab:statistics} shows some general statistics about the FAIR Jupyter KG, which consists of about 190~million triples taking up a total of about 20.6~GB in space. 
Taking inspiration from the Scholia frontend to Wikidata 
\cite{NielsenF2017Scholia},
we anticipate using the KG for creating profiles based on specific entity types. 
To this end,
we created the FAIR Jupyter KG from multiple smaller graphs based on separate mappings for each entity type. 
We present the time it took to construct the KG, the number of mapping rules retrieved, the total triples generated for each entity and the total file size of the RDF file generated.


\begin{table}[!htb]
\caption{General statistics of the FAIR Jupyter knowledge graph. The graphs for four entity types (CellName, CodeAnalysis, RepositoryFile, and MarkdownFeature) were omitted from the prototype implementation for performance reasons. 
}
\label{tab:statistics}
\begin{tabular}{| p{0.30\linewidth} | r | r | r |r |}
\hline
\textbf{Entity} & \textbf{Time (in sec)} & \textbf{No. of mappings}  & \textbf{Triples generated} & \textbf{File Size}\\
\hline
Article & 5.2 & 17 & 60589 & 7.8 MB\\
Author & 5.1 & 6 & 121247 & 14.1 MB\\
Cell & 39.8 & 10 & 5940797 & 645.9 MB\\
CellFeature & 15.4 & 11 & 1340610 &  168.4 MB\\
CellModule & 8.4 & 2 & 917367 & 120.2 MB\\
CellName  & 176.2 & 12 & 38609232 & 4.2 GB\\
CellExecution & 8.4 & 15 & 125383 & 48.9 MB\\
CodeAnalysis & 462.4 & 162  & 77295103 & 7.9 GB\\
Journal & 4.7 & 7 & 4497 & 0.49 MB\\
MarkdownFeature & 203.5 & 125 & 36693762 & 3,8 GB\\
MESH & 43.2 & 3 & 9078 & 1.2 MB\\
Notebook & 10.5 & 24 & 627127 & 65.2 MB\\
NotebookAST & 32.0 & 159 & 3281939 & 327.7 MB\\
NotebookCodeStyle & 14.5 & 4 & 216105 & 26.4 MB\\
NotebookFeature & 8.5 & 29 & 374071 & 42.3 MB\\
NotebookMarkdown & 37.5 & 125 & 3389551 & 353.2 MB\\
NotebookModule & 9.4 & 31 & 498504 & 63.3 MB\\
NotebookName & 47.4 & 115 & 619562 & 137.2 MB\\
Repository & 7.1 & 38 & 183592 & 20.1 MB\\
RepositoryFile  & 100.6 & 5 & 19223736 & 2.6 GB \\
RepositoryRelease & 6.4 & 9 & 252500 & 33.3 MB \\
RequirementFile & 5.5 & 6 & 27865 & 14.3 MB\\
\hline
Total & 1251.7 & 915 & 189812217 &  20.6 GB \\ 
\hline

\end{tabular}
\end{table}



\begin{sidewaysfigure}
\includegraphics[width=\textwidth]{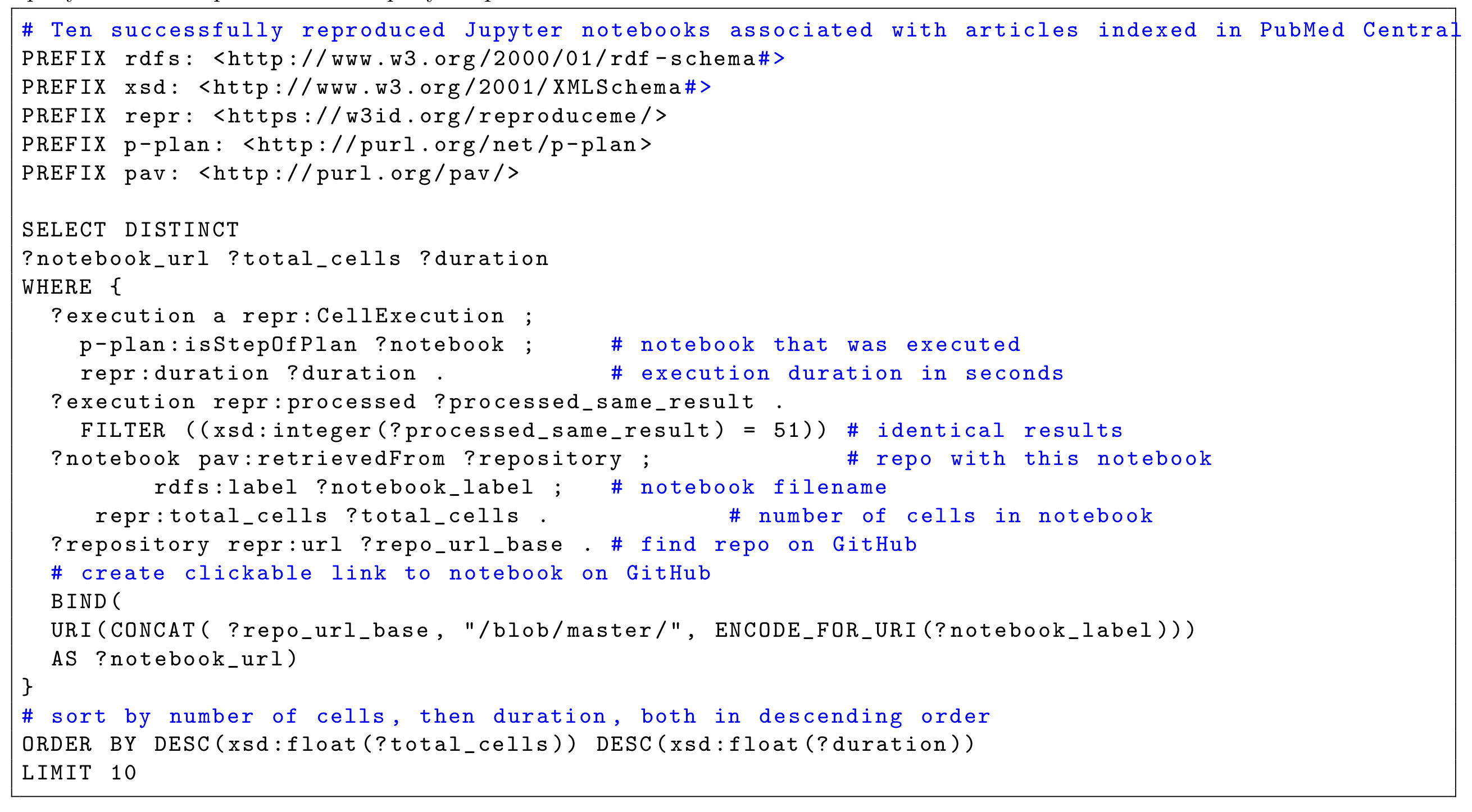}
\centering
\caption{Example query: Ten successfully reproduced Jupyter notebooks associated with articles indexed in PubMed Central. The query text can be pasted into the query endpoint as is.} 
\label{lst:tennotebooks}
\end{sidewaysfigure}

\newpage

With the architecture laid out in 
Figure 
\ref{fig:model}
and
Table \ref{tab:statistics},  
it becomes possible to interrogate the dataset about any of the entities, classes or their relationships in a granular fashion. 
An example query is provided in Figure
\ref{lst:tennotebooks}, which asks for notebooks where the reproducibility run produced results identical to those reported in the original publication. 
It then sorts this set of notebooks by the number of cells and the execution duration of the notebook (both of these parameters can serve as a proxy for the complexity of the notebook itself or the computations triggered by it), and then it limits the results to 10.

In the following, we illustrate the diversity of possible queries by presenting three sets of further example queries.
First, Table \ref{tab:reproduce} shows queries corresponding to some figures 
and tables from
the publication describing the original dataset \cite{samuel2024computational}. 
Second, Table \ref{tab:otherqueries} shows a brief selection of other queries that can be queried over the FAIR Jupyter graph.
Third, federated queries can be run that combine information from our knowledge graph with other knowledge graph~-- some examples of federated queries with Wikidata are given in
Table \ref{tab:federation}.
A more comprehensive list of queries is available at \url{https://w3id.org/fairjupyter}.

Multiple queries with results involving the same entity types can be combined into a profile for that entity type, as described in \cite{NielsenF2017Scholia}.

\begin{table}
\centering
\caption{SPARQL queries to the knowledge graph that reproduce materials from the original manuscript describing the dataset \cite{samuel2024computational}
}
\label{tab:reproduce}
\begin{tabular}{|p{0.20\linewidth} | p{0.8\linewidth} |}
\hline
\textbf{Figure no. in  \cite{samuel2024computational}} & \textbf{SPARQL query} \\
\hline
Fig. 3 & \href{https://reproduceme.uni-jena.de/#/dataset/fairjupyter/query?query=SELECT%20%3Fresearch_field%20%28COUNT%28DISTINCT%20%3Farticle%29%20AS%20%3Fnumber_of_articles%29%0AWHERE%20%7B%20%20%0A%20%20%3Frepository%20%3Chttp%3A%2F%2Fpurl.org%2Fpav%2FretrievedFrom%3E%20%3Farticle%20.%0A%20%20%3Farticle%20%3Chttp%3A%2F%2Fwww.w3.org%2Fns%2Fprov-o%23specializationOf%3E%20%3Fmesh%20.%0A%20%20%3Fmesh%20%3Chttp%3A%2F%2Fwww.w3.org%2Fns%2Fprov-o%23generalizationOf%3E%20%3Ftop_mesh%20.%0A%20%20%3Ftop_mesh%20%3Chttp%3A%2F%2Fwww.w3.org%2F2000%2F01%2Frdf-schema%23label%3E%20%3Fresearch_field%0A%20%20%0A%7D%0AGROUP%20BY%20%3Fresearch_field%0AORDER%20BY%20DESC%28%3Fnumber_of_articles%29%0ALIMIT%2010%0A}{Research articles by research field} \\ 
\hline
Fig. 4 & \href{https://reproduceme.uni-jena.de/#/dataset/fairjupyter/query?query=PREFIX%20rdfs%3A%20%3Chttp%3A%2F%2Fwww.w3.org%2F2000%2F01%2Frdf-schema%23%3E%0APREFIX%20xsd%3A%20%3Chttp%3A%2F%2Fwww.w3.org%2F2001%2FXMLSchema%23%3E%0APREFIX%20repr%3A%20%3Chttps%3A%2F%2Fw3id.org%2Freproduceme%2F%3E%0A%0ASELECT%20%3Fresearch_field%20%28COUNT%28DISTINCT%20%3Frepository%29%20as%20%3Frepository_count%29%20%28COUNT%28DISTINCT%20%3Frepository_nb%29%20as%20%3Frepositories_with_notebooks_count%29%0AWHERE%20%7B%0A%20%20%7B%0A%20%20%3Frepository%20%3Chttp%3A%2F%2Fpurl.org%2Fpav%2FretrievedFrom%3E%20%3Farticle%20.%0A%20%20%7D%0A%20%20UNION%0A%20%20%7B%0A%20%20%3Frepository_nb%20%3Chttp%3A%2F%2Fpurl.org%2Fpav%2FretrievedFrom%3E%20%20%3Farticle%20%3B%0A%20%20%20%20%20%20%20%20%20%20%20%20%20%20repr%3Anotebooks_count%20%3Fnotebooks_count%20.%0A%20%20%09FILTER%28xsd%3Ainteger%28%3Fnotebooks_count%29%20%3E%200%29%0A%20%20%7D%0A%20%20%3Farticle%20%3Chttp%3A%2F%2Fwww.w3.org%2Fns%2Fprov-o%23specializationOf%3E%20%3Fmesh%20.%0A%20%20%3Fmesh%20%3Chttp%3A%2F%2Fwww.w3.org%2Fns%2Fprov-o%23generalizationOf%3E%20%3Ftop_mesh%20.%0A%20%20%3Ftop_mesh%20rdfs%3Alabel%20%3Fresearch_field%0A%7D%0AGROUP%20BY%20%3Fresearch_field%0AORDER%20BY%20DESC%28%3Frepository_count%29%0ALIMIT%2010}{Research field (MeSH terms) by the number of GitHub repositories  that contain at least one Jupyter notebook.} \\ 
\hline
Fig. 5 & \href{https://reproduceme.uni-jena.de/#/dataset/fairjupyter/query?query=SELECT%20%3Fjournal_name%20%28COUNT%28%3Farticle%29%20as%20%3Farticle_count%29%0AWHERE%20%7B%0A%20%20%3Farticle%20%3Chttps%3A%2F%2Fw3id.org%2Freproduceme%2FpublishedIn%3E%20%3Fjournal%20.%0A%20%20%3Fjournal%20%3Chttp%3A%2F%2Fwww.w3.org%2F2000%2F01%2Frdf-schema%23label%3E%20%3Fjournal_name%20.%0A%7D%0AGROUP%20BY%20%3Fjournal_name%0AORDER%20BY%20DESC%28%3Farticle_count%29%0ALIMIT%2010}{Journals with the highest number of articles that had a valid GitHub repository and at least one Jupyter notebook.} \\ 
\hline
Fig. 6 & \href{https://reproduceme.uni-jena.de/#/dataset/fairjupyter/query?query=SELECT%20%3Fjournal_name%20%28COUNT%28%3Frepository%29%20as%20%3Frepository_count%29%20%28COUNT%28%3Frepository_nb%29%20as%20%3Frepositories_with_notebooks_count%29%20WHERE%0A%20%20%7B%0A%20%20%3Farticle%20%3Chttps%3A%2F%2Fw3id.org%2Freproduceme%2FpublishedIn%3E%20%3Fjournal%20.%0A%20%20%3Fjournal%20%3Chttp%3A%2F%2Fwww.w3.org%2F2000%2F01%2Frdf-schema%23label%3E%20%3Fjournal_name%20.%0A%20%20%7B%0A%20%20%3Frepository%20%3Chttp%3A%2F%2Fpurl.org%2Fpav%2FretrievedFrom%3E%20%3Farticle%20.%0A%20%20%7D%0A%20%20UNION%0A%20%20%7B%0A%20%20%3Frepository_nb%20%3Chttp%3A%2F%2Fpurl.org%2Fpav%2FretrievedFrom%3E%20%3Farticle%20%3B%0A%20%20%20%20%20%20%20%20%20%20%20%20%20%20%3Chttps%3A%2F%2Fw3id.org%2Freproduceme%2Fnotebooks_count%3E%20%3Fnotebooks_count%20.%0A%20%20FILTER%28%3Chttp%3A%2F%2Fwww.w3.org%2F2001%2FXMLSchema%23integer%3E%28%3Fnotebooks_count%29%20%3E%200%29%0A%20%20%7D%0A%20%20%7D%0A%20%20GROUP%20BY%20%3Fjournal_name%0A%0AORDER%20BY%20DESC%28%3Frepository_count%29%0ALIMIT%2010}{Journals by the number of GitHub repositories and by the number of GitHub repositories with at least one Jupyter notebook.} \\ 
\hline
Fig. 7 & \href{https://reproduceme.uni-jena.de/#/dataset/fairjupyter/query?query=SELECT%20%3Fjournal_name%20%28COUNT%28%3Frepository_nb%29%20AS%20%3Frepositories_with_notebooks_count%29%0A%20%20%20%20%20%20%20%20%3Fmax_notebooks_count%0AWHERE%20%7B%0A%20%20%7B%0A%20%20%20%20SELECT%20%3Fjournal%20%28MAX%28%3Fnotebooks_count%29%20AS%20%3Fmax_notebooks_count%29%0A%20%20%20%20WHERE%20%7B%0A%20%20%20%20%20%20%3Farticle%20%3Chttps%3A%2F%2Fw3id.org%2Freproduceme%2FpublishedIn%3E%20%3Fjournal%20.%0A%20%20%20%20%20%20%3Fjournal%20%3Chttp%3A%2F%2Fwww.w3.org%2F2000%2F01%2Frdf-schema%23label%3E%20%3Fjournal_name%20.%20%20%0A%20%20%20%20%20%20%3Frepository_nb%20%3Chttp%3A%2F%2Fpurl.org%2Fpav%2FretrievedFrom%3E%20%3Farticle%20%3B%0A%20%20%20%20%20%20%20%20%20%20%20%20%20%20%20%20%20%20%20%20%20%3Chttps%3A%2F%2Fw3id.org%2Freproduceme%2Fnotebooks_count%3E%20%3Fnotebooks_count%20.%0A%20%20%20%20%20%20FILTER%28%3Chttp%3A%2F%2Fwww.w3.org%2F2001%2FXMLSchema%23integer%3E%28%3Fnotebooks_count%29%20%3E%200%29%20%20%0A%20%20%20%20%7D%0A%20%20%20%20GROUP%20BY%20%3Fjournal%0A%20%20%7D%0A%20%20%3Farticle%20%3Chttps%3A%2F%2Fw3id.org%2Freproduceme%2FpublishedIn%3E%20%3Fjournal%20.%0A%20%20%3Fjournal%20%3Chttp%3A%2F%2Fwww.w3.org%2F2000%2F01%2Frdf-schema%23label%3E%20%3Fjournal_name%20.%20%20%0A%20%20%3Frepository_nb%20%3Chttp%3A%2F%2Fpurl.org%2Fpav%2FretrievedFrom%3E%20%3Farticle%20%3B%0A%20%20%20%20%20%20%20%20%20%20%20%20%20%20%20%20%20%3Chttps%3A%2F%2Fw3id.org%2Freproduceme%2Fnotebooks_count%3E%20%3Fnotebooks_count%20.%0A%20%20FILTER%28%3Chttp%3A%2F%2Fwww.w3.org%2F2001%2FXMLSchema%23integer%3E%28%3Fnotebooks_count%29%20%3E%200%29%20%20%0A%7D%0AGROUP%20BY%20%3Fjournal_name%20%20%3Fmax_notebooks_count%0AORDER%20BY%20DESC%28%3Frepositories_with_notebooks_count%29%0ALIMIT%2010%0A}{Journals by number of GitHub repositories with Jupyter notebooks.} \\ 
\hline
Fig. 9 & \href{https://reproduceme.uni-jena.de/#/dataset/fairjupyter/query?query=SELECT%20%3Flanguage%20%28COUNT%28%3Fnotebook%29%20as%20%3Fnotebook_count%29%0AWHERE%20%7B%0A%20%20%3Fnotebook%20a%20%3Chttps%3A%2F%2Fw3id.org%2Freproduceme%2FNotebook%3E%20%3B%0A%20%20%20%20%20%20%20%20%20%20%20%20%3Chttps%3A%2F%2Fw3id.org%2Freproduceme%2Flanguage%3E%20%3Flanguage%20.%0A%7D%0AGROUP%20BY%20%3Flanguage%0AORDER%20BY%20DESC%28%3Fnotebook_count%29%0ALIMIT%2010}{Programming languages of the notebooks.} \\ 
\hline
Fig. 10 & \href{https://reproduceme.uni-jena.de/#/dataset/fairjupyter/query?query=SELECT%20%3Fcreated_year%20%3Flanguage%20%28COUNT%28%3Fnotebook%29%20as%20%3Fnotebook_count%29%0AWHERE%20%7B%0A%20%20%3Fnotebook%20a%20%3Chttps%3A%2F%2Fw3id.org%2Freproduceme%2FNotebook%3E%20%3B%0A%20%20%20%20%20%20%20%20%20%20%20%20%3Chttp%3A%2F%2Fpurl.org%2Fpav%2FretrievedFrom%3E%20%20%3Frepository%20%3B%0A%20%20%20%20%20%20%20%20%20%20%20%20%3Chttps%3A%2F%2Fw3id.org%2Freproduceme%2Flanguage%3E%20%3Flanguage%20%3B%0A%20%20%20%20%20%20%20%20%20%20%20%20%3Chttps%3A%2F%2Fw3id.org%2Freproduceme%2Flanguage_version%3E%20%3Fversion%20.%0A%20%20%3Frepository%20%3Chttps%3A%2F%2Fw3id.org%2Freproduceme%2Fcreated_at%3E%20%3Fcreated_at%20.%0A%20%20BIND%28REPLACE%28str%28%3Fcreated_at%29%2C%20%22%28%5C%5Cd%2A%29-.%2A%22%2C%20%22%241%22%29%20AS%20%3Fcreated_year%29%20%20%0A%7D%0AGROUP%20BY%20%3Fcreated_year%20%3Flanguage%0AORDER%20BY%20%3Fcreated_year%20%3Flanguage%0A%0A}{Relative proportion of the most frequent programming languages used in the notebooks per year.} \\ 
\hline
Fig. 11 & \href{https://reproduceme.uni-jena.de/#/dataset/fairjupyter/query?query=SELECT%20%3Fcreated_year%20%3Fminor_version%20%28COUNT%28%3Fnotebook%29%20as%20%3Fcount_minor_version%29%0AWHERE%20%7B%0A%20%20%3Fnotebook%20a%20%3Chttps%3A%2F%2Fw3id.org%2Freproduceme%2FNotebook%3E%20%3B%0A%20%20%20%20%20%20%20%20%20%20%20%20%3Chttp%3A%2F%2Fpurl.org%2Fpav%2FretrievedFrom%3E%20%20%3Frepository%20%3B%0A%20%20%20%20%20%20%20%20%20%20%20%20%3Chttps%3A%2F%2Fw3id.org%2Freproduceme%2Flanguage%3E%20%22python%22%20%3B%0A%20%20%20%20%20%20%20%20%20%20%20%20%3Chttps%3A%2F%2Fw3id.org%2Freproduceme%2Flanguage_version%3E%20%3Fversion%20.%0A%20%20%3Frepository%20%3Chttps%3A%2F%2Fw3id.org%2Freproduceme%2Fcreated_at%3E%20%3Fcreated_at%20.%0A%20%20BIND%28REPLACE%28str%28%3Fcreated_at%29%2C%20%22%28%5C%5Cd%2A%29-.%2A%22%2C%20%22%241%22%29%20AS%20%3Fcreated_year%29%20%20%0A%20%20BIND%28SUBSTR%28%3Fversion%2C%201%2C%203%29%20AS%20%3Fminor_version%29%0A%20%20FILTER%28%3Fversion%20%21%3D%20%223%22%20%26%26%20%3Fversion%20%21%3D%20%221%22%20%26%26%20%3Fversion%20%21%3D%20%22ES2015%22%29%0A%7D%0AGROUP%20BY%20%3Fcreated_year%20%3Fminor_version%0AORDER%20BY%20%3Fcreated_year%20%3Fminor_version%0A%0A}{Python notebooks by minor Python version by year of last commit to the GitHub repository containing the notebook.} \\ 
\hline
Fig. 12 & \href{https://reproduceme.uni-jena.de/#/dataset/fairjupyter/query?query=SELECT%20%3Fcreated_year%20%3Fmajor_version%20%28COUNT%28%3Fnotebook%29%20as%20%3Fcount_major_version%29%0AWHERE%20%7B%0A%20%20%3Fnotebook%20a%20%3Chttps%3A%2F%2Fw3id.org%2Freproduceme%2FNotebook%3E%20%3B%0A%20%20%20%20%20%20%20%20%20%20%20%20%3Chttp%3A%2F%2Fpurl.org%2Fpav%2FretrievedFrom%3E%20%20%3Frepository%20%3B%0A%20%20%20%20%20%20%20%20%20%20%20%20%3Chttps%3A%2F%2Fw3id.org%2Freproduceme%2Flanguage%3E%20%22python%22%20%3B%0A%20%20%20%20%20%20%20%20%20%20%20%20%3Chttps%3A%2F%2Fw3id.org%2Freproduceme%2Flanguage_version%3E%20%3Fversion%20.%0A%20%20%3Frepository%20%3Chttps%3A%2F%2Fw3id.org%2Freproduceme%2Fcreated_at%3E%20%3Fcreated_at%20.%0A%20%20BIND%28REPLACE%28str%28%3Fcreated_at%29%2C%20%22%28%5C%5Cd%2A%29-.%2A%22%2C%20%22%241%22%29%20AS%20%3Fcreated_year%29%20%20%0A%20%20BIND%28SUBSTR%28%3Fversion%2C%201%2C%201%29%20AS%20%3Fmajor_version%29%0A%20%20FILTER%28%3Fversion%20%21%3D%20%223%22%20%26%26%20%3Fversion%20%21%3D%20%221%22%20%26%26%20%3Fversion%20%21%3D%20%22ES2015%22%29%0A%7D%0AGROUP%20BY%20%3Fcreated_year%20%3Fmajor_version%0AORDER%20BY%20%3Fcreated_year%20%3Fmajor_version%0A%0A}{Python notebooks by major Python version by year of first commit to the notebook’s GitHub repository.} 
\\ 
\hline
Fig. 19 & \href{https://reproduceme.uni-jena.de/#/dataset/fairjupyter/query?query=SELECT%20%3Fexception%20%28COUNT%28%3Fexception%29%20AS%20%3Fcount%29%0AWHERE%20%7B%0A%20%20%3Fexecution%20%20a%20%3Chttps%3A%2F%2Fw3id.org%2Freproduceme%2FCellExecution%3E%20%3B%0A%20%20%20%20%3Chttps%3A%2F%2Fw3id.org%2Freproduceme%2Fexception%3E%20%3Fexception%20.%0A%7D%0AGROUP%20BY%20%3Fexception%0AORDER%20BY%20DESC%28%3Fcount%29%0ALIMIT%2010%0A}{Exceptions occurring in Jupyter notebooks in our corpus.} \\ 
\hline
Fig. 22 & \href{https://reproduceme.uni-jena.de/#/dataset/fairjupyter/query?query=PREFIX%20xsd%3A%20%3Chttp%3A%2F%2Fwww.w3.org%2F2001%2FXMLSchema%23%3E%0ASELECT%20DISTINCT%20%3Fresearch_field%20%28COUNT%28%3Fexception%29%20AS%20%3Fexception_count%29%0AWHERE%20%7B%20%20%0A%20%20%3Fexecution%20%20a%20%3Chttps%3A%2F%2Fw3id.org%2Freproduceme%2FCellExecution%3E%20%3B%0A%20%20%20%20%3Chttps%3A%2F%2Fw3id.org%2Freproduceme%2Fexception%3E%20%3Fexception%20%3B%0A%20%20%20%20%3Chttp%3A%2F%2Fpurl.org%2Fpav%2FretrievedFrom%3E%20%3Frepository%20.%0A%20%20%3Frepository%20a%20%3Chttp%3A%2F%2Fusefulinc.com%2Fns%2Fdoap%23GitRepository%3E%20%3B%0A%20%20%09%09%09%3Chttp%3A%2F%2Fpurl.org%2Fpav%2FretrievedFrom%3E%20%3Farticle%20%3B%0A%20%20%09%09%09%3Chttps%3A%2F%2Fw3id.org%2Freproduceme%2Fnotebooks_count%3E%20%3Fnotebooks_count%20.%0A%20%20%3Farticle%20a%20%3Chttp%3A%2F%2Fpurl.org%2Fspar%2Ffabio%2FArticle%3E%20%3B%20%0A%20%20%09%09%20%3Chttp%3A%2F%2Fwww.w3.org%2Fns%2Fprov-o%23specializationOf%3E%20%3Fmesh%20.%0A%20%20%3Fmesh%20%3Chttp%3A%2F%2Fwww.w3.org%2Fns%2Fprov-o%23generalizationOf%3E%20%3Ftop_mesh%20.%0A%20%20%3Ftop_mesh%20%3Chttp%3A%2F%2Fwww.w3.org%2F2000%2F01%2Frdf-schema%23label%3E%20%3Fresearch_field%20.%20%20%20%20%0A%20%20FILTER%20%28xsd%3Ainteger%28%3Fnotebooks_count%29%3E0%29%0A%7D%0AGROUP%20BY%20%3Fresearch_field%0AORDER%20BY%20DESC%28%3Fexception_count%29%0ALIMIT%2010%0A}{Jupyter notebook exceptions by research field, taking as a proxy the highest-level MeSH terms of the article associated with the notebook.} \\ 
\hline
Table 2 & \href{https://reproduceme.uni-jena.de/#/dataset/fairjupyter/query?query=PREFIX%20xsd%3A%20%3Chttp%3A%2F%2Fwww.w3.org%2F2001%2FXMLSchema%23%3E%0ASELECT%20%28COUNT%28%3Fprocessed_different_result%29%20AS%20%3Fcount_different_result%29%20%28COUNT%28%3Fprocessed_same_result%29%20AS%20%3Fcount_same_result%29%20%28%3Fcount_same_result%20%2B%20%3Fcount_different_result%20AS%20%3Fcount_successful_executions%29%0AWHERE%20%7B%0A%20%20%3Fexecution%20a%20%3Chttps%3A%2F%2Fw3id.org%2Freproduceme%2FCellExecution%3E%20.%0A%20%20OPTIONAL%20%7B%20%3Fexecution%20%3Chttps%3A%2F%2Fw3id.org%2Freproduceme%2Fexception%3E%20%3Fexception%20.%20%7D%0A%20%20OPTIONAL%20%7B%0A%20%20%20%20%3Fexecution%20%3Chttps%3A%2F%2Fw3id.org%2Freproduceme%2Fprocessed%3E%20%3Fprocessed_different_result%20.%0A%20%20%20%20FILTER%20%28%28xsd%3Ainteger%28%3Fprocessed_different_result%29%20%3D%2035%29%20%26%26%20%21bound%28%3Fexception%29%29%0A%20%20%7D%0A%20%20OPTIONAL%20%7B%0A%20%20%20%20%3Fexecution%20%3Chttps%3A%2F%2Fw3id.org%2Freproduceme%2Fprocessed%3E%20%3Fprocessed_same_result%20.%0A%20%20%20%20FILTER%20%28%28xsd%3Ainteger%28%3Fprocessed_same_result%29%20%3D%2051%29%20%26%26%20%21bound%28%3Fexception%29%29%0A%20%20%7D%0A%20%20%0A%20%20%0A%7D}{Notebooks with successful executions with same and different results} \\ 
\hline
Table 4 & \href{https://reproduceme.uni-jena.de/#/dataset/fairjupyter/query?query=SELECT%20%3Fnotebook%20%3Ferror%20%3Fdescription%0AWHERE%20%7B%0A%20%20%3Ferror%20a%20%3Chttps%3A%2F%2Fw3id.org%2Freproduceme%2FNotebookCodeStyleError%3E%20%3B%0A%20%20%20%20%20%20%20%20%3Chttps%3A%2F%2Fw3id.org%2Freproduceme%2Fdescription%3E%20%3Fdescription%20%3B%0A%20%20%20%20%20%20%20%20%3Chttp%3A%2F%2Fpurl.org%2Fpav%2FretrievedFrom%3E%20%3Fnotebook%20.%0A%7D}{Common Python code warnings/ style errors in our notebook corpus.} \\
\hline
\end{tabular}
\end{table}

\begin{table}
\centering
\caption{Other queries over the FAIR Jupyter graph}
\label{tab:otherqueries}
\begin{tabular}{|p{0.8\linewidth} | p{0.2\linewidth} |}
\hline
\textbf{Description} & \textbf{SPARQL query} \\
\hline
Notebooks by search term: `immun' AND (`stem' OR `differentiation') & \href{https://reproduceme.uni-jena.de/#/dataset/fairjupyter/query?query=SELECT%20DISTINCT%20%3Fnotebook_url%20%3Farticle_label%20%3Fkeywords%20WHERE%20%7B%20%0A%20%20%3Farticle%20%3Chttps%3A%2F%2Fw3id.org%2Freproduceme%2Fkeywords%3E%20%3Fkeywords%20.%0A%20%20%3Farticle%20%3Chttp%3A%2F%2Fwww.w3.org%2F2000%2F01%2Frdf-schema%23label%3E%20%3Farticle_label%20.%20%0A%20%20%3Farticle%20%3Chttps%3A%2F%2Fw3id.org%2Freproduceme%2FpublishedIn%3E%20%3Fjournal%20.%0A%20%20%3Fjournal%20%3Chttp%3A%2F%2Fwww.w3.org%2F2000%2F01%2Frdf-schema%23label%3E%20%3Fjournal_label%20.%20%0A%20%20FILTER%20%28REGEX%28LCASE%28CONCAT%28%3Fkeywords%2C%20%22%20%22%2C%20%3Farticle_label%2C%20%22%20%22%2C%20%3Fjournal_label%29%29%2C%20%22immun%22%29%29%0A%20%20FILTER%20%28REGEX%28LCASE%28CONCAT%28%3Fkeywords%2C%20%22%20%22%2C%20%3Farticle_label%2C%20%22%20%22%2C%20%3Fjournal_label%29%29%2C%20%22%5C%5Cb%28stem%7Cdifferentiation%29%22%29%29%0A%20%20%3Farticle%20%5E%3Chttp%3A%2F%2Fpurl.org%2Fpav%2FretrievedFrom%3E%20%3Frepository%20.%0A%20%20%3Fnotebook%20%3Chttp%3A%2F%2Fpurl.org%2Fpav%2FretrievedFrom%3E%20%3Frepository%20.%0A%20%20%3Fnotebook%20%3Chttp%3A%2F%2Fwww.w3.org%2F1999%2F02%2F22-rdf-syntax-ns%23type%3E%20%3Chttps%3A%2F%2Fw3id.org%2Freproduceme%2FNotebook%3E%20.%0A%20%20%3Fnotebook%20%3Chttp%3A%2F%2Fwww.w3.org%2F2000%2F01%2Frdf-schema%23label%3E%20%3Fnotebook_label%20.%20%23%20filename%0A%20%20%3Frepository%20%3Chttps%3A%2F%2Fw3id.org%2Freproduceme%2Furl%3E%20%3Frepo_url_base%20.%20%23%20find%20repo%20on%20GitHub%0A%20%20BIND%28URI%28CONCAT%28%20%3Frepo_url_base%2C%20%22%2Fblob%2Fmaster%2F%22%2C%20%3Fnotebook_label%29%29%20AS%20%3Fnotebook_url%29%20%23%20create%20clickable%20link%20to%20notebook%20on%20GitHub%0A%20%20FILTER%20%28%3Fnotebook_url%20%21%3D%20%22%22%29%0A%7D%0A}{Query} \\
\hline
Article by keywords, e.g., `open source' & \href{https://reproduceme.uni-jena.de/#/dataset/fairjupyter/query?query=SELECT%20DISTINCT%20%3Farticle%20%3Fkeywords%20WHERE%20%7B%20%0A%20%20%3Farticle%20%3Chttps%3A%2F%2Fw3id.org%2Freproduceme%2Fkeywords%3E%20%3Fkeywords%20.%0A%20%20FILTER%20%28REGEX%28LCASE%28%3Fkeywords%29%2C%20%22open%28.%29source%22%29%29%0A%7D%0A}{Query} \\
\hline
Most common errors in immunology & \href{https://reproduceme.uni-jena.de/#/dataset/fairjupyter/query?query=SELECT%20DISTINCT%20%3Fexception%20%28COUNT%28%3Fexception%29%20AS%20%3Fcount%29%20WHERE%20%7B%20%0A%20%20%20%20%3Fexecution%20%20a%20%3Chttps%3A%2F%2Fw3id.org%2Freproduceme%2FCellExecution%3E%20%3B%0A%20%20%20%20%3Chttps%3A%2F%2Fw3id.org%2Freproduceme%2Fexception%3E%20%3Fexception%20%3B%0A%20%20%20%20%3Chttp%3A%2F%2Fpurl.org%2Fpav%2FretrievedFrom%3E%20%3Frepository%20.%0A%20%20%20%20%3Frepository%20a%20%3Chttp%3A%2F%2Fusefulinc.com%2Fns%2Fdoap%23GitRepository%3E%20%3B%0A%20%20%09%09%09%3Chttp%3A%2F%2Fpurl.org%2Fpav%2FretrievedFrom%3E%20%3Farticle%20.%0A%20%20%3Farticle%20%20%3Chttps%3A%2F%2Fw3id.org%2Freproduceme%2Fkeywords%3E%20%3Fkeywords%20.%0A%20%20FILTER%20%28REGEX%28LCASE%28%3Fkeywords%29%2C%20%22immun%22%29%29%0A%7D%0AGROUP%20BY%20%3Fexception%0AORDER%20BY%20DESC%28%3Fcount%29%0A}{Query}\\
\hline
Most common errors in Nature journal & \href{https://reproduceme.uni-jena.de/#/dataset/fairjupyter/query?query=PREFIX%20rdfs%3A%20%3Chttp%3A%2F%2Fwww.w3.org%2F2000%2F01%2Frdf-schema%23%3E%0ASELECT%20%20%3Fexception%20%28COUNT%28%3Fexception%29%20AS%20%3Fcount%29%20WHERE%20%7B%20%0A%20%20%20%20%3Fexecution%20%20a%20%3Chttps%3A%2F%2Fw3id.org%2Freproduceme%2FCellExecution%3E%20%3B%0A%20%20%20%20%3Chttps%3A%2F%2Fw3id.org%2Freproduceme%2Fexception%3E%20%3Fexception%20%3B%0A%20%20%20%20%3Chttp%3A%2F%2Fpurl.org%2Fpav%2FretrievedFrom%3E%20%3Frepository%20.%0A%20%20%20%20%3Frepository%20a%20%3Chttp%3A%2F%2Fusefulinc.com%2Fns%2Fdoap%23GitRepository%3E%20%3B%0A%20%20%09%09%09%3Chttp%3A%2F%2Fpurl.org%2Fpav%2FretrievedFrom%3E%20%3Farticle%20.%0A%20%20%3Farticle%20%20%3Chttps%3A%2F%2Fw3id.org%2Freproduceme%2FpublishedIn%3E%20%3Fjournal%20.%0A%20%20%3Fjournal%20rdfs%3Alabel%20%3Fjournal_name%0A%20%20FILTER%20%28%3Fjournal_name%3D%22Nature%22%29%0A%7D%0AGROUP%20BY%20%3Fexception%0AORDER%20BY%20DESC%28%3Fcount%29%0A}{Query}\\
\hline
MeSH terms ranked by `ModuleNotFoundError' frequency & \href{https://reproduceme.uni-jena.de/#/dataset/fairjupyter/query?query=PREFIX%20xsd%3A%20%3Chttp%3A%2F%2Fwww.w3.org%2F2001%2FXMLSchema%23%3E%0ASELECT%20DISTINCT%20%3Fresearch_field%20%28COUNT%28%3Fexception%29%20AS%20%3Fexception_count%29%0AWHERE%20%7B%20%20%0A%20%20%3Fexecution%20%20a%20%3Chttps%3A%2F%2Fw3id.org%2Freproduceme%2FCellExecution%3E%20%3B%0A%20%20%20%20%3Chttps%3A%2F%2Fw3id.org%2Freproduceme%2Fexception%3E%20%3Fexception%20%3B%0A%20%20%20%20%3Chttp%3A%2F%2Fpurl.org%2Fpav%2FretrievedFrom%3E%20%3Frepository%20.%0A%20%20%3Frepository%20a%20%3Chttp%3A%2F%2Fusefulinc.com%2Fns%2Fdoap%23GitRepository%3E%20%3B%0A%20%20%09%09%09%3Chttp%3A%2F%2Fpurl.org%2Fpav%2FretrievedFrom%3E%20%3Farticle%20%3B%0A%20%20%09%09%09%3Chttps%3A%2F%2Fw3id.org%2Freproduceme%2Fnotebooks_count%3E%20%3Fnotebooks_count%20.%0A%20%20%3Farticle%20a%20%3Chttp%3A%2F%2Fpurl.org%2Fspar%2Ffabio%2FArticle%3E%20%3B%20%0A%20%20%09%09%20%3Chttp%3A%2F%2Fwww.w3.org%2Fns%2Fprov-o%23specializationOf%3E%20%3Fmesh%20.%0A%20%20%3Fmesh%20%3Chttp%3A%2F%2Fwww.w3.org%2Fns%2Fprov-o%23generalizationOf%3E%20%3Ftop_mesh%20.%0A%20%20%3Ftop_mesh%20%3Chttp%3A%2F%2Fwww.w3.org%2F2000%2F01%2Frdf-schema%23label%3E%20%3Fresearch_field%20.%20%20%20%20%0A%20%20FILTER%20%28%3Fexception%3D%27ModuleNotFoundError%27%29%0A%7D%0AGROUP%20BY%20%3Fresearch_field%0AORDER%20BY%20DESC%28%3Fexception_count%29}{Query} \\
\hline
Repositories by their stargazers count & \href{https://reproduceme.uni-jena.de/#/dataset/fairjupyter/query?query=PREFIX%20xsd%3A%20%3Chttp%3A%2F%2Fwww.w3.org%2F2001%2FXMLSchema%23%3E%0ASELECT%20DISTINCT%20%3Frepo%20%3Fstargazers_count%20WHERE%20%7B%0A%20%20%3Frepo%20%3Chttps%3A%2F%2Fw3id.org%2Freproduceme%2Fstargazers_count%3E%20%3Fcount.%20%0A%20%20BIND%28xsd%3Afloat%28%3Fcount%29%20AS%20%3Fstargazers_count%29%0A%20%20FILTER%20%28%28%3Fstargazers_count%29%20%3E%200%29%0A%7D%20%0AORDER%20BY%20DESC%28%3Fstargazers_count%29}{Query}\\
\hline
\end{tabular}
\end{table}


\begin{table}
\centering
\caption{Federated queries between the FAIR Jupyter KG and Wikidata}
\label{tab:federation}
\begin{tabular}{|p{0.8\linewidth} | p{0.2\linewidth} |}
\hline
\textbf{Description} & \textbf{SPARQL query} \\
\hline
 Match articles between FAIR Jupyter and Wikidata via DOI & \href{https://reproduceme.uni-jena.de/#/dataset/fairjupyter/query?query=PREFIX%20rdfs%3A%20%3Chttp%3A%2F%2Fwww.w3.org%2F2000%2F01%2Frdf-schema%23%3E%0A%0APREFIX%20wikidata_wd%3A%20%3Chttp%3A%2F%2Fwww.wikidata.org%2Fentity%2F%3E%0APREFIX%20wikidata_wdt%3A%20%3Chttp%3A%2F%2Fwww.wikidata.org%2Fprop%2Fdirect%2F%3E%0A%0ASELECT%20DISTINCT%0A%0A%20%3Ffj_article%0A%20%3Fwikidata%0A%20%3Fwikidata_label%0A%20%3FDOI%0A%0AWHERE%20%7B%0A%20%20%3Ffj_article%20%3Chttps%3A%2F%2Fw3id.org%2Freproduceme%2Fdoi%3E%20%3Fdoi%20.%0A%20%20BIND%28UCASE%28%3Fdoi%29%20AS%20%3FDOI%29%0A%20%20service%20%3Chttps%3A%2F%2Fquery.wikidata.org%2Fsparql%3E%20%7B%0A%20%20%20%20%3Fwikidata%20wikidata_wdt%3AP356%20%3FDOI%20.%0A%20%20%20%20%3Fwikidata%20rdfs%3Alabel%20%3Fwikidata_label%20.%0A%20%20%20%20FILTER%20%28LANG%28%3Fwikidata_label%29%20%3D%20%22en%22%29%0A%20%20%7D%0A%7D%0ALIMIT%20100}{Query} \\
\hline
Match articles between FAIR Jupyter and Wikidata via PMC ID & \href{https://reproduceme.uni-jena.de/#/dataset/fairjupyter/query?query=PREFIX%20rdfs%3A%20%3Chttp%3A%2F%2Fwww.w3.org%2F2000%2F01%2Frdf-schema%23%3E%0A%0APREFIX%20wikidata_wd%3A%20%3Chttp%3A%2F%2Fwww.wikidata.org%2Fentity%2F%3E%0APREFIX%20wikidata_wdt%3A%20%3Chttp%3A%2F%2Fwww.wikidata.org%2Fprop%2Fdirect%2F%3E%0A%0ASELECT%20DISTINCT%0A%0A%20%3Ffj_article%0A%20%3Fwikidata%0A%20%3Fwikidata_label%0A%20%3Fpmc%0A%0AWHERE%20%7B%0A%20%20%3Ffj_article%20%3Chttps%3A%2F%2Fw3id.org%2Freproduceme%2Fpmc%3E%20%3Fpmc%20.%0A%20%20service%20%3Chttps%3A%2F%2Fquery.wikidata.org%2Fsparql%3E%20%7B%0A%20%20%20%20%3Fwikidata%20wikidata_wdt%3AP932%20%3Fpmc%20.%0A%20%20%20%20%3Fwikidata%20rdfs%3Alabel%20%3Fwikidata_label%20.%0A%20%20%20%20FILTER%20(LANG(%3Fwikidata_label)%20%3D%20%22en%22)%0A%20%20%7D%0A%7D%0ALIMIT%20100}{Query} \\
\hline
Match articles between FAIR Jupyter and Wikidata via MeSH in different language, i.e Malayalam & \href{https://reproduceme.uni-jena.de/#/dataset/fairjupyter/query?query=PREFIX%20rdfs%3A%20%3Chttp%3A%2F%2Fwww.w3.org%2F2000%2F01%2Frdf-schema%23%3E%0A%0APREFIX%20wikidata_wd%3A%20%3Chttp%3A%2F%2Fwww.wikidata.org%2Fentity%2F%3E%0APREFIX%20wikidata_wdt%3A%20%3Chttp%3A%2F%2Fwww.wikidata.org%2Fprop%2Fdirect%2F%3E%0A%0ASELECT%20DISTINCT%0A%0A%20%3Ffj_article%0A%20%3Fwikidata%0A%20%3Fwikidata_label%0A%20%3FDOI%0A%0AWHERE%20%7B%0A%20%20%3Ffj_article%20%3Chttp%3A%2F%2Fwww.w3.org%2Fns%2Fprov-o%23specializationOf%3E%20%3Fmesh_url%20.%0A%20%20BIND%28REPLACE%28STR%28%3Fmesh_url%29%2C%20%22.%2AMESH%2FD%22%2C%20%22D%22%29%20AS%20%3FMESH%29%20%0A%20%20service%20%3Chttps%3A%2F%2Fquery.wikidata.org%2Fsparql%3E%20%7B%0A%20%20%20%20%3Fwikidata%20wikidata_wdt%3AP486%20%3FMESH%20.%0A%20%20%20%20%3Fwikidata%20rdfs%3Alabel%20%3Fwikidata_label%20.%0A%20%20%20%20FILTER%20%28LANG%28%3Fwikidata_label%29%20%3D%20%22ml%22%29%0A%20%20%7D%0A%7D%0ALIMIT%20100}{Query} \\
\hline
\end{tabular}
\end{table}





\section{Discussion}
\label{sec:discussion}

In this work, we have demonstrated how the accessibility of a shared dataset can be improved by leveraging 
semantic web approaches
and representing it in a knowledge graph that can be explored from a web browser.

That dataset had been created using a workflow for reproducing Jupyter notebooks from biomedical publications, and in the present work, we have enhanced it by representing the dataset's components~-- publications, GitHub repositories, Jupyter notebooks and reproducibility aspects thereof~-- using web ontologies.

The dataset is of particular interest for trainings and education as well as for showcasing real-world examples of actual research practices, from which both best practices and things to avoid can be synthesized. 
%
In order to enhance the dataset's usefulness in such contexts, we have paid special attention to 
its 
Findability, Accessiblity, Interoperability and Reusability, as per 
the 
FAIR Principles \cite{wilkinson2016fair}
for sharing 
research data. Since our data is about software, we also took into account adaptations of the FAIR Principles to software \cite{lamprecht2019Towards}.

The original dataset as a whole was already well aligned with the FAIR Principles,
and we added an additional layer of alignment by sharing  individual elements of the original dataset in a FAIR way, 
so as to make it easier for humans and machines to engage with them. 
At a very basic 
level, 
the original dataset as a whole 
becomes more findable by virtue of
the current manuscript and the FAIR Jupyter website pointing to it, more accessible by being available in additional formats (e.g.\ RDF),
more interoperable by compatibility with additional standards (particularly ontologies, as per Figure \ref{fig:model}) 
and more reusable through combinations of the above as well as the enhanced granularity. 
At a more profound level, 
the dataset's individual facets and features~-- be it 
the classes shown in 
Figure \ref{fig:model}
or
the entities listed in 
Table \ref{tab:statistics}
or any of the relationships between them~--
have become more FAIR, as they
can now be searched, explored, aggregated, filtered
and reused in additional user friendly ways both individually or in various combinations, both manually and programmatically.
This 
knowledge graph approach
facilitates additional routes for engaging with the original data and
makes it easy to
ask and address questions that were not included in the paper describing the original dataset.
It also
 provides an additional level of reproducibility in that it allows for reproducing specific aspects of the original paper, as demonstrated with 
Table
\ref{tab:reproduce}. 
Finally, the coupling of the knowledge graph with a public-facing web frontend 
and user-friendly example queries essentially 
turns the dataset from FAIR data into a FAIR service.

Enhancing a dataset this way and setting up such a knowledge graph service represents an additional effort in terms of data sharing. We cannot say at this point whether those efforts~-- which we have tried to outline in this manuscript~-- merit the actual benefits in our case, yet our example queries outline some of the potential benefits, and we plan to assess usage at a later stage and to  tailor future enhancements accordingly.
We are welcoming others to experiment with our workflows or to collaborate with us to adapt them to their use cases.

Assuming a baseline level of usefulness of a service like FAIR Jupyter, keeping it useful requires updates in a timely manner. When we ran 
our Jupyter reproducibility pipeline that resulted in the original dataset, 
our search query in PubMed Central (cf.\ 
\ref{foot:search-query})
yielded 3467 articles in March 2023, as opposed to 1419 in February 2021 and 
5126 in April 2024.
We are working on establishing a pipeline that would regularly feed new articles and their notebooks into our knowledge graph and on highlighting the successful reproductions, e.g. by badges \cite{kidwell2016badges}
displayed next to them or through nanopublications 
\cite{bucur2023nanopublication}
sharing reproducibility information about them. 

We are working towards extending the knowledge graph in other ways, too.
For instance, 
we are exploring to include~-- or to federate~-- information about institutions and citations pertaining to articles with Jupyter notebooks, perhaps together with information about resources used alongside Jupyter (e.g. as per Figure 29 in \cite{samuel2024computational}). 
We are also conscious that
our way of using MeSH terms~-- which are assigned by PubMed at an article-level~-- does not necessarily represent an optimal way to associate topics with notebooks, for which approaches like a BERTopic \cite{grootendorst2022bertopic}
analysis of the notebooks themselves might be more suitable.

Furthermore, we could combine FAIR Jupyter with a ReproduceMeGit \cite{samuel2021reproducemegit} service that would
take 
Jupyter notebooks
as input, 
assess their reproducibility   (potentially even before the corresponding article is submitted to a publication venue) and~-- depending on the outcome~-- invoke the knowledge graph to suggest similar notebooks (e.g. based on dependencies or topics) with better reproducibility, better documentation or fewer styling issues as a mechanism to help notebook authors to familiarize themselves with best practices.

The reproducibility of any particular notebook is not set in stone, and research questions could be formed around that (e.g.\ which dependencies contribute most to reproducibility decay, and how that changes over time), which an updated knowledge graph could help address, perhaps seeded by incorporating data from our initial run of the pipeline in 2021. 
%
Other lines of research could look at deep dependencies \cite{nesbitt2024biomedical}
of the Jupyter notebooks or their repositories and relate these to aspects of reproducibility~-- as discussed here~-- or of security, e.g.\ as per \cite{lins2024critical}.
%
As our pipeline is  automated, it lends itself to
further 
integration with other automated workflows, for example in the context of workflow systems (e.g.\ \cite{goble2020fair}) or machine-actionable data management plans (e.g.\ \cite{Miksa2019Ten}).


In its current state, FAIR Jupyter can already support various uses in educational settings, be that the identification of articles to choose for reprohack events
\cite{hettne2020ReprohackNL}, materials for lessons by The Carpentries \cite{pugachev2019carpentries}, for general reproducibility activities in libraries \cite{granger2020How}
or for self-guided study by anyone wishing to learn about Jupyter, Python, software dependencies or computational reproducibility.

We welcome
contributions from others~-- including from other disciplines (e.g.\ as per \cite{wofford2020Jupyter})~-- to the reproducibility pipeline, to the knowledge graph or to any of their suggested improvements or applications.




\section{Resource Availability Statement}
\label{sec:resource}
The FAIR Jupyter service with links to the SPARQL endpoint, code and all the corresponding resources is available at \url{https://w3id.org/fairjupyter} with GPL-3.0 license. 
We are committed to keeping it up for five years, i.e.\ until April 2029.
The code used for generating the original dataset is available at \url{https://github.com/fusion-jena/computational-reproducibility-pmc}. The csv files, the YARRRML and RML mappings used for constructing the knowledge graph are available at \url{https://github.com/fusion-jena/fairjupyter}.
We will provide Zenodo snapshots of the GitHub code when submitting the camera-ready version of the manuscript.

\subsection{Environmental footprint}
To estimate the environmental impact of our computation, we leveraged a tool from the Green Algorithms project (http://www.green-algorithms.org/). This tool calculates the environmental footprint based on hardware configuration, total runtime, and location.
To generate the original dataset, the pipeline consumed 373.78 kWh, resulting in a carbon footprint of approximately 126.58 kg CO2e, equivalent to 11.51 tree years when using default values for Germany. The pipeline for constructing the KG took 20.8 minutes, resulting in a carbon footprint of 7.33 g CO2e. The carbon footprint of the query execution from Table \ref{tab:reproduce} is around 151.48mg CO2e.

\subsection{Ethical considerations}
The original dataset contained the email addresses of the corresponding authors, which are available from PubMed Central as part of the respective article's full text. We have not included these author email addresses within the publicly available knowledge graph, 
since the increased accessibility of the data in the graph also increases the potential for misuse. However, we retain these emails internally to facilitate communication with authors regarding their repositories and the reproducibility of their work.

%
%
%
\bibliographystyle{splncs04}
\bibliography{main}
\end{document}